# The dangers of performance-based research funding in non-competitive higher education systems[1]

*Giovanni Abramo[a,b,\*], Tindaro Cicero[a], Ciriaco Andrea D'Angelo[a],*

[a] Laboratory for Studies of Research and Technology Transfer
School of Engineering, Department of Management
University of Rome "Tor Vergata"

[b] National Research Council of Italy

**Abstract**

An increasing number of nations allocate public funds to research institutions on the basis of rankings obtained from national evaluation exercises. Therefore, in non-competitive higher education systems where top scientists are dispersed among all the universities, rather than concentrated among a few, there is a high risk of penalizing those top scientists who work in lower-performance universities. Using a five-year bibliometric analysis conducted on all Italian universities active in the hard sciences from 2004-2008, this work analyzes the distribution of publications and relevant citations by scientists within the universities, measures the research performance of individual scientists, quantifies the intensity of concentration of top scientists at each university, provides performance rankings for the universities, and indicates the effects of selective funding on the top scientists of low-ranked universities.

[1] Abramo, G., Cicero, T., D'Angelo, C.A. (2011). The dangers of performance-based research funding in non-competitive higher education systems. *Scientometrics*, 87(3), 641-654. DOI: 10.1007/s11192-011-0355-4

# 1. Introduction

In recent years, an increasing number of nations, are carrying out national research evaluation exercises, with one of their aims being to improve allocative efficiency in public funding of individual institutions. Governments and their national agencies are gradually imposing elements of competition (Georghiou and Larédo, 2005) mainly based on: i) evaluations of project proposals; ii) implementation of performance-based research funding (PBRF) systems. Examples of this second type of system are the various Research Assessment Exercises of the United Kingdom (from 1986 to 2008), which have paved the way to the upcoming Research Evaluation Framework (to be completed in 2014), and also Australia's Excellence in Research initiative (launched in 2010), which follows the earlier Research Quantum Composite Index (since 1995), New Zealand's Performance-Based Research Fund (2006), and in Italy, the first Triennial Research Evaluation (2006), soon to be followed by a Quinquennial Research Evaluation (expected in 2011).

Scholars, policy makers and administrators of research institutions are ever more involved in debates over the value of such exercises and, more generally, on the advisability or not of implementing PBRF incentive systems (Orr et al., 2007; Strehl et al., 2007; Shattock, 2004; Debackere and Glänzel, 2004; Rousseau and Smeyers, 2000). Geuna and Martin (2003) provide an exhaustive analysis of advantages and disadvantages of PBRF. Specific studies have been directed at questions regarding possible adverse effects of the incentive schemes on researcher behavior (Bhattacharya and Newhouse, 2008; Moed, 2008; Langford et al., 2006; Laudel, 2006; Butler, 2003). Some scholars have directly examined whether researchers adapt their personal production objectives to the evaluation criteria. Butler (2003), examining the Australian research system, and Moed (2008), for the English system, show that researchers tend to align themselves with the evaluation exercise guidelines. Auranen and Nieminen (2010), comparing funding environments for university research in eight countries, come to the opposite conclusion: that no straightforward connection exists between financial incentives and the efficiency of university systems. There seems to be a growing consensus for the desirability of permanent adoption of these types of initiative given a primary objective of stimulating and rewarding excellence in research institutions. However, the desired macroeconomic aims of such systems can only be obtained if internal redistribution of government resources within each research institution follows a consistent logic: the intended effects of national evaluation systems can result only if a "funds for quality" rule is followed at all levels of decision-making. If this does not occur, then the objective of stimulating greater efficiency may not be reached, since there will be insufficient incentives at the level of individual scientists. This becomes more likely with increasing dispersion of performance within the research institutions. For the Italian case, as demonstrated by Abramo et al. (2010)[2], the overall research product of each university is significantly concentrated in the output of a small number of scientists. Italian universities all show a particularly skewed distribution of performance, which makes them all very similar. The variability of performance between universities is lower than within. With a scenario of this type, without a mechanism for internal redistribution coherent with that adopted by the central government for allocating portions of research funding to universities, the objectives of the PBRF system will not be realized. National evaluation exercises do not provide the universities with performance rankings at the level of individual scientists, and the universities lack suitable instruments for their measure. As a consequence, in higher education systems where performance differences within universities are notably higher than between, government funding allocations based on university rankings rather than on individual scientist rankings are likely to fall short of their purpose.

---
[2]This manuscript is currently under consideration for publication; an abstract is available at http://www.disp.uniroma2.it/laboratorioRTT/TESTI/Working%20paper/RESPOL_Cicero.pdf, last access on January 26, 2011.

Our objective is to provide evidence of the risks of selective funding allocations at institutional level in those higher education systems characterized by high dispersion of performance within universities. We will show the possible penalization that top scientists in low-ranked universities may experience, in funding terms. Through a bibliometric analysis conducted on all Italian universities active in the hard sciences for the five years from 2004-2008, we will analyze the distribution of performance within universities in two scenarios: under the observed scenario with their top scientists and in a hypothetical scenario "without". We will measure the concentrations of top scientists in each university and relate these to the performance classifications for the universities. Simulating a selective allocation of resources to universities, similar to that generally adopted, we will show what effects such a system has on top scientists of low-ranked universities, and thus on the macroeconomic objective of rewarding excellence.

The next section of the work presents the methodology used, the dataset for the analyses, and the type of indicator used to measure both individual and university performance. Section 3 illustrates the results of the analyses conducted and the last section presents the related policy implications.

## 2. Methodology

The field of observation consists of all Italian universities active in the hard sciences (a total of 77 institutions), observed over the five-year period 2004-2008. Research performance is measured through bibliometric techniques, essentially applying an impact indicator related to citations. For the hard sciences, the literature gives ample justification for this choice of performance indicator suggesting that : i) scientific publications are a good proxy of overall research output (Moed et al., 2004); and ii) citations are a good proxy of impact on scientific advancement, notwithstanding the possible distortions inherent in this indicator (Glanzel, 2008).

### 2.1 Dataset

The dataset used for the analyses was constructed based on the Observatory of Public Research (ORP)[3], a database developed and maintained by the authors under license from the Thomson Reuters Web of Science (WoS). Beginning from the Italian raw data indexed in the WoS, then applying a complex algorithm for reconciliation of the authors' affiliation and disambiguation of the true identity of the authors, each publication (articles, reviews, and conference proceedings) is attributed to the university scientists that produced it, with an error of less than 5% (D'Angelo et al., 2011).

---
[3] www.orp.researchvalue.it, last access on January 26, 2011.

In the Italian system of classification, the hard sciences are grouped in nine University Disciplinary Areas (UDAs[4]) and 205 Scientific Disciplinary Sector (SDSs[5]). Each university researcher is classified in only one SDS[6]. To render the analyses more significant, the field of observation was limited to those SDSs in which at least 50% of the scientists produced at least one publication in the period 2004-2008. In the 183 SDSs thus examined, over the 2004-2008 period, there were an average of 39,508 scientists (equal to 58.3% of total Italian university researchers), distributed in 77 universities (Table 1).

| UDA | N. of SDSs | Universities | Research staff |
|---|---|---|---|
| Mathematics and computer sciences | 9 | 64 | 3,515 |
| Physics | 8 | 61 | 2,873 |
| Chemistry | 11 | 59 | 3,603 |
| Earth sciences | 12 | 48 | 1,439 |
| Biology | 19 | 66 | 5,785 |
| Medicine | 47 | 55 | 12,196 |
| Agricultural and veterinary sciences | 28 | 48 | 3,153 |
| Civil engineering | 7 | 49 | 1,455 |
| Industrial and information engineering | 42 | 68 | 5,489 |
| **Total** | **183** | **77** | **39,508** |

*Table 1: Universities and research staff in the hard sciences of the Italian academic system; data 2004-2008*

### 2.2 Indicators

For measurement of research performance, the study uses the indicator Scientific Strength (*SS*), equal to the sum of the standardized citations of all publications by a researcher, each standardized citation divided by the number of co-authors of the publication. Citations of a publication are standardized dividing them by the median[7] of citations[8] of all Italian publications of the same year and WoS subject category[9].

---

[4] Mathematics and computer sciences; Physics; Chemistry; Earth sciences; Biology; Medicine; Agriculture and veterinary sciences; Civil engineering; Industrial and information engineering.

[5] Complete list available at http://cercauniversita.cineca.it/php5/settori/index.php, last access on January 26, 2011.

[6] In Italy, all personnel enter the university system through public examinations, and career advancement also requires such public examinations. Classification of scientists into SDSs, make it easier to formulate examinations calls and select members of the examination committees.

[7] The decision to standardize citations with respect to the median number (rather than to the average, as frequently observed in the literature) is justified by the fact that the distribution of citations is highly skewed (Lundberg, 2007).

[8] Observed at 30/06/2009.

[9] When a publication falls in two or more subject categories the average of the medians is used.

In the case of the life sciences (biology, biomedical research, and clinical medicine)[10] different weights have been given to each co-author according to his/her position in the list and the character of the co-authorship (intra-mural or extra-mural). For each scientist in the dataset, the indicator described is subsequently divided by the number of years out of the five observed in which the individual held an official post in an Italian university faculty. The performance of each scientist is thus compared with that of his/her colleagues in the same SDS, in order to take into account the varying intensity of publication and citation for different sectors (Abramo and D'Angelo, 2011). Expressing individual performance within each SDS as a percentile rank, it is then possible to compare scientists belonging to different disciplines.

The university performance rankings are prepared: i) by SDS, calculating the above indicator through aggregation of the standardized citations for the researchers that compose it and dividing by the number of research staff in the SDS; ii) by UDA, aggregating the data for the SDSs through operations of standardization and weighting, to avoid possible distortion inherent in analyses at this level of aggregation. The analytic formula for the Scientific Strength ($SS_{UDA}$) in a given UDA of a given university is:

$$SS_{UDA} = \sum_{s=1}^{n_{UDA}} \left( \frac{SS_s}{P_s^*} \cdot \frac{Add_s}{Add_{UDA}} \right)$$

where:
$SS_s$ = SS per faculty-member of the SDS $s$
$P_s^*$ = average SS of Italian universities in SDS $s$
$Add_s$ = number of scientist in SDS $s$
$Add_{UDA}$ = number of scientists in the UDA
$n_{UDA}$ = number of SDSs in the UDA

The operations for standardizing and weighting values for the SDSs of a UDA permit taking into account the diverse fertility of the SDSs and their varying representativeness, in terms of members, in each UDA (Abramo et al., 2008).

## 3. Results

### 3.1 Distribution of performance for Italian university researchers

Our earlier work shows that the Italian higher education system has high or very high levels of concentration of performance within universities (Abramo et al., 2010). This is caused in part by a substantial share of non-productive researchers[11]. In the 183 SDSs analyzed, there are 6,640 "non-productives" out of total 39,508 scientists (17%). The descriptive statistics for the non-productives, by UDA, are presented in Table 2. The UDA showing the lowest share of non-productives is Chemistry (7%), with a minimum for the SDS of Organic chemistry (CHIM/06), at 4%. The UDA Civil engineering shows the highest percentage (29%). It should be noted that the percentages of non-productives are inversely correlated to the average intensity of publication in the various sectors. On average, for all the SDSs, 29% of the researchers produce 71% of the total output.

---
[10] In these fields order in the authors' list reflects the varying contribution of the authors to the article. In other fields the alphabetical order is the norm.
[11] "Non-productive" researchers are defined as those for whom it is not possible to identify articles, reviews or conference proceedings in the WoS, for the period 2004-2008.

| UDA | N. of SDSs | Min | Max | Average |
|---|---|---|---|---|
| Mathematics and computer sciences | 9 | 11% (MAT/09) | 37% (MAT/01) | 23% |
| Physics | 8 | 7% (FIS/03) | 49% (FIS/08) | 16% |
| Chemistry | 11 | 4% (CHIM/06) | 13% (CHIM/12) | 7% |
| Earth sciences | 12 | 10% (GEO/03) | 42% (GEO/05) | 19% |
| Biology | 19 | 5% (BIO/18) | 34% (BIO/03) | 14% |
| Medicine | 47 | 6% (MED/07) | 43% (MED/45) | 20% |
| Agricultural and veterinary sciences | 28 | 5% (VET/03) | 50% (AGR/10) | 18% |
| Civil engineering | 7 | 22% (ICAR/03) | 40% (ICAR/05) | 29% |
| Industrial and information engineering | 42 | 2% (ING-IND/34) | 48% (ING-IND/02) | 21% |

*Table 2: Descriptive statistics for percentages of non-productive researchers, by UDA*

Next we observe that out of 39,508 researchers, 9,701 (25% of total) have a nil impact[12]. Table 3 presents the descriptive statistics for these researchers. The UDA that registers the lowest percentage of scientists with nil SS is again Chemistry (9% on average), with an absolute minimum in CHIM/06 (5%). The highest percentage is seen in Civil engineering (47% on average), although the absolute maximum for an SDS is seen in Industrial and Information engineering, specifically in Naval and Marine construction and installation (ING-IND/02). In general, there is strong heterogeneity of data for the SDSs within the individual areas. For example, Industrial and Information engineering, which we observed has the SDS with the absolute maximum percentage of scientists at nil SS, also shows a minimum value as low as 7% (Industrial bioengineering - ING-IND/34). In Physics, the figures for minimum (8%) and maximum (68%) of scientists with nil impact again show a remarkable spread, at least between the two extremes, for History of physics, FIS/08 and Physics of condensed matter, FIS/03.

| UDA | N. of SDSs | Min | Max | Average |
|---|---|---|---|---|
| Mathematics and computer sciences | 9 | 18% (MAT/09) | 53% (MAT/03) | 37% |
| Physics | 8 | 8% (FIS/03) | 68% (FIS/08) | 23% |
| Chemistry | 11 | 5% (CHIM/03) | 15% (CHIM/12) | 9% |
| Earth sciences | 12 | 12% (GEO/10) | 59% (GEO/05) | 27% |
| Biology | 19 | 9% (BIO/17) | 43% (BIO/02) | 18% |
| Medicine | 47 | 8% (MED/07) | 50% (MED/35 - 45) | 25% |
| Agricultural and veterinary sciences | 28 | 9% (VET/02) | 78% (AGR/10) | 31% |
| Civil engineering | 7 | 35% (ICAR/03) | 62% (ICAR/05) | 47% |
| Industrial and Information engineering | 42 | 7% (ING-IND/34) | 87% (ING-IND/02) | 36% |

*Table 3: Descriptive statistics for percentages of researchers with nil impact, by UDA*

Overall, 23% of researchers produce 77% of the total scientific strength. Naturally, this has a great influence on measures of inequality: Table 4 presents the ratio between cumulative performance of scientists situated in the bottom 40% and the top 20%[13] of each SDS rank listing.

---

[12] In this analysis, the scientists with nil SS include the non-productives and the researchers who published but did not receive any citations.

| UDA | N. of SDSs | Max | Average |
|---|---|---|---|
| Mathematics and computer sciences | 9 | 0.064 (MAT/09) | 0.013 |
| Physics | 8 | 0.106 (FIS/03) | 0.044 |
| Chemistry | 11 | 0.234 (CHIM/11) | 0.105 |
| Earth sciences | 12 | 0.125 (GEO/10) | 0.036 |
| Biology | 19 | 0.129 (BIO/19) | 0.041 |
| Medicine | 47 | 0.077 (MED/03) | 0.015 |
| Agricultural and veterinary sciences | 28 | 0.101 (VET/02) | 0.026 |
| Civil engineering | 7 | 0.010 (ICAR/03) | 0.002 |
| Industrial and information engineering | 42 | 0.137 (ING-IND/24) | 0.025 |

*Table 4: Descriptive statistics for the ratio of cumulative SS from the bottom 40% to cumulative SS from top 20% of scientists, by UDA*

The average values of the ratio confirm the high values of concentration within each disciplinary area. Chemistry is somewhat of an exception, and in Chemistry and biotechnology of fermentations - CHIM/11, the ratio reaches an absolute maximum value (0.234). The absolute minimum occurs in an SDS of Civil engineering: in Environmental and Health Engineering (ICAR/03), where the bottom 40% cumulatively produce 1% of cumulative output from the top 20%.

### 3.2 The impact of top scientists on university rankings

Allocation of resources to universities on the basis of the rankings obtained from national research assessment exercise can reduce or totally deprive some universities of public funds. Further, universities generally do not posses the instruments for comparative evaluation of single scientist performance. Considering that research groups have various capabilities in negotiating funding, not necessarily related to merits, this could then cause internal allocation of resources that is not efficient. Given this situation, it is interesting to ask what would happen to the ranking position of each university if the top scientists were excluded from the evaluation. In this section we analyze the observed performances of universities as well as their performances under a hypothetical case in which bibliometric analysis excludes the top scientists, equal in number to 20% of the research staff of each SDS at a university. As an example, we present the situation of universities operating in CHIM/06. There are 35 universities with a staff of at least five researchers. Table 5 shows their relative rankings under the two scenarios, the positive or negative shift and the Gini coeffecient[14] calculated for the observed scenario of distribution of all research staff in the SDS at the university. In this scenario, we can observe immediately that high-ranked universities have high Gini coefficient. The two sets of rankings related to observed and hypothetical scenario are significantly correlated (Spearman $\rho = 0.655$, p-value $< 0.000$), though in the presence of considerable shifts in ranking in many universities. Out of a total of 35 universities, 19 gain position, 12 lose position and four remain stable. Of the 19 universities that gain position, eight rise more than five positions; while of the 12 universities that decline, six descend more than five positions. We also observe that in the cases of the greater negative shifts, there is also a greater value of concentration, and vice versa. For example, the University of Trieste, which loses 26 positions between the first and second scenarios, has a concentration value of 0.783. Vice versa, the University of Rome "Tor Vergata", with a concentration value for performance of 0.415, is the university that registers the greatest jump in rank (+15 positions). In general, we observe greater positive shifts in position for universities with lesser concentration of performance.

---

[13] This measure varies between 0 and 100%. The closer the ratio to zero, the greater the difference in average performance between these sub-populations, noting also that the first group (bottom 40%) is always double the number of the second (top 20%).

[14] Gini coefficient here is a measure of the inequality of research productivity: a value of 0 suggests that the variation among scientists is nil: a value of 1 indicates maximal inequality.

To evaluate the extent of the phenomenon we carried out a correlation analysis between variation in ranking and the Gini coefficient. Figure 1 presents the dispersion graph and the trend line for data concerning the 35 universities considered. The Spearman value, equaling -0.805, is highly significant (p-value $< 0.000$), indicating that the universities with more homogenous performance in the SDS rise in position between the observed scenario and the analysis that excludes the top performers in the SDS, and vice versa. The same correlation calculations as for CHIM/06 were carried out for all the SDSs in the field of observation, limiting the analysis, for reasons of significance, to SDSs with at least 5 universities[15]. The results, presented in aggregate form per UDA, in Table 6, consistently show negative correlation values, except in a very few cases (GEO/03, VET/10, ING-INF/07), which were all not statistically significant.

| Universities | Observed scenario (rank) | Hypothetical scenario (rank) | Sign | Δ | Gini* |
|---|---|---|---|---|---|
| University of Trieste | 1 | 27 | - | 26 | 0.783 |
| University of L'Aquila | 2 | 2 | = | 0 | 0.630 |
| University of Pisa | 3 | 13 | - | 10 | 0.667 |
| University of Ferrara | 4 | 31 | - | 27 | 0.730 |
| University of Padova | 5 | 1 | + | 4 | 0.549 |
| University of Bologna | 6 | 11 | - | 5 | 0.607 |
| University of Salerno | 7 | 7 | = | 0 | 0.525 |
| University of Camerino | 8 | 4 | + | 4 | 0.536 |
| University of Palermo | 9 | 3 | + | 6 | 0.487 |
| University of Siena | 10 | 14 | - | 4 | 0.499 |
| University of Parma | 11 | 8 | + | 3 | 0.479 |
| University of Florence | 12 | 12 | = | 0 | 0.531 |
| University of Basilicata | 13 | 5 | + | 8 | 0.384 |
| University of Sassari | 14 | 16 | - | 2 | 0.550 |
| University of Rome "La Sapienza" | 15 | 17 | - | 2 | 0.572 |
| University of Eastern Piedmont | 16 | 22 | - | 6 | 0.593 |
| University of Perugia | 17 | 9 | + | 8 | 0.403 |
| University of Pavia | 18 | 25 | - | 7 | 0.610 |

---

[15] With less than five observations the Gini coefficient would lose significance.

| University of Calabria | 19 | 15 | + | 4 | 0.509 |
| University of Insubria | 20 | 10 | + | 10 | 0.358 |
| University of Rome "Tor Vergata" | 21 | 6 | + | 15 | 0.415 |
| University of Turin | 22 | 26 | - | 4 | 0.561 |
| University of Venice "Ca' Foscari" | 23 | 21 | + | 2 | 0.461 |
| University of Urbino "Carlo Bo" | 24 | 20 | + | 4 | 0.445 |
| University of Naples "Federico II" | 25 | 18 | + | 7 | 0.449 |
| University of Milan | 26 | 24 | + | 2 | 0.521 |
| University of Bari | 27 | 23 | + | 4 | 0.536 |
| University of Genova | 28 | 34 | - | 6 | 0.641 |
| University of Milan Bicocca | 29 | 19 | + | 10 | 0.251 |
| University of Messina | 30 | 29 | + | 1 | 0.394 |
| Politechnic of Ancona | 31 | 30 | + | 1 | 0.385 |
| University of Catania | 32 | 33 | - | 1 | 0.480 |
| University of Modena and Reggio E. | 33 | 32 | + | 1 | 0.460 |
| University of Chieti "G. D'Annunzio" | 34 | 28 | + | 6 | 0.244 |
| University of Cagliari | 35 | 35 | = | 0 | 0.520 |

*Table 5: Variations in performance ranking under observed and hypothetical scenarios for universities in Organic chemistry, with values for concentration*
[*]*Data for the observed scenario*

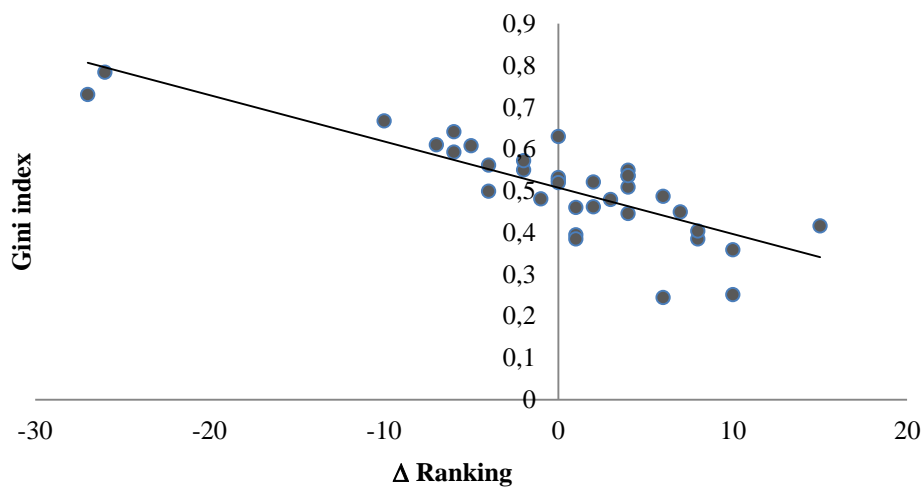

*Figure 1: Correlation between variations in ranking (between observed and hypothetical scenarios) and Gini index for the 35 universities active in CHIM/06*

| UDA | Min | Max | Significant SDSs[*] |
|---|---|---|---|
| Mathematics and computer sciences | -0.883 (MAT/06[**]) | -0.432 (MAT/03[**]) | 8 out of 8 |
| Physics | -0.853 (FIS/03[**]) | -0.561 (FIS/02[**]) | 6 out of 7 |
| Chemistry | -0.838 (CHIM/02[**]) | -0.435 (CHIM/09[*]) | 7 out of 8 |
| Earth sciences | -0.909 (GEO/10[*]) | 0.055 (GEO/03) | 6 out of 10 |
| Biology | -0.896 (BIO/19[**]) | -0.221 (BIO/08) | 14 out of 18 |
| Medicine | -0.956 (MED/46[**]) | -0.110 (MED/12) | 28 out of 39 |
| Agricultural and veterinary sciences | -0.746 (AGR/12[**]) | 0.109 (VET/10) | 10 out of 25 |
| Civil engineering | -0.699 (ICAR/08[**]) | -0.522 (ICAR/09[**]) | 5 out of 6 |
| Industrial and Information engineering | -0.986 (ING-IND/06[**]) | 0.053 (ING-INF/07) | 13 out of 28 |

*Table 6: Descriptive statistics for correlations between variations in ranking (between observed and hypothetical scenarios) and Gini index for the SDSs in each UDA*
*Significance level: \* p-value < 0.05; \*\* p-value < 0.01*

The same simulation as for CHIM/06 was conducted for all the SDSs, to provide university rankings at the level of UDA.

The statistics for comparison between rankings are presented in Table 7.

| UDA | N of Variations | Max Δ - | Max Δ + | Average | Median |
|---|---|---|---|---|---|
| Mathematics and computer sciences | 42 out of 50 (84%) | -11 | 10 | 2.7 | 3.0 |
| Physics | 46 out of 48 (96%) | -17 | 16 | 5.8 | 4.0 |
| Chemistry | 34 out of 41 (83%) | -8 | 10 | 2.7 | 2.0 |
| Earth sciences | 29 out of 32 (91%) | -8 | 5 | 2.3 | 2.0 |
| Biology | 44 out of 49 (90%) | -15 | 13 | 4.5 | 4.0 |
| Medicine | 36 out of 42 (86%) | -11 | 11 | 4.0 | 3.0 |
| Agricultural and veterinary sciences | 22 out of 25 (88%) | -6 | 6 | 2.5 | 2.0 |
| Civil engineering | 24 out of 31 (77%) | -11 | 10 | 2.7 | 2.0 |
| Industrial and information engineering | 40 out of 42 (95%) | -21 | 22 | 6.1 | 4.0 |

*Table 7: Descriptive statistics for shifts in performance ranking of universities, for each UDA*

In general, we see that the percentage of universities affected by variation in rank ranges from 77% in Civil engineering to 96% in Physics. Industrial and information engineering shows both the maximum negative (-21) and positive (+22) shifts between the two rankings. In the other areas the maximum shifts, while lesser, are still sizeable, especially in Physics (+16), Biology (+13), Medicine (+11), as well as in Mathematics and computer sciences, Chemistry and Civil engineering (+10). There are notable negative shifts in the areas of Physics (-17), Biology (-15), and Mathematics and computer science, Medicine and Civil engineering (both -11).

For an example in detail, we show the situation for the Industrial and information engineering UDA, which registers the greatest variability between the two sets of rankings. Over the period considered, there are 42 Italian universities with at least 5 scientists in at least one of the SDSs in this area. Table 8 shows the shifts in quintile between the university rankings obtained from the observed distribution and those from the hypothetical scenario, eliminating the top scientists.

| | | Hypothetical scenario | | | | | |
|---|---|---|---|---|---|---|---|
| | Performance | Very High | High | Medium | Low | Very Low | Total |
| Observed | Very high | 7 | 2 | 0 | 0 | 0 | 9 |
| | High | 2 | 2 | 1 | 2 | 1 | 8 |
| | Medium | 0 | 2 | 4 | 1 | 1 | 8 |
| | Low | 0 | 2 | 2 | 4 | 0 | 8 |
| | Very Low | 0 | 0 | 1 | 1 | 7 | 9 |
| | Total | 9 | 8 | 8 | 8 | 9 | 42 |

*Table 8: Variations in performance (quintile) of universities active in Industrial and information engineering, between observed and hypothetical scenarios*

Out of 42 universities, 24 (57% of total) do not show any variation in quintile. Of the remaining 18 universities, 11 shift to an adjacent quintile: four to the next lowest and seven to the next highest. Another six universities shift two quintiles: three shift downwards and three upwards. The University of Bari registers the greatest shift, moving downwards a full 3 quintiles: without its top scientists, the performance of this university drops from "High" to "Very Low".

### 3.4 The distribution of top national scientists

In the preceding sections we showed that with the high levels of concentration such as seen in the Italian universities, if top scientists of each university are excluded from performance calculations, there are substantial changes and reversals in university rankings.

In this section we assess the extent to which research funding systems based on university rankings can penalize the nation's best scientists. We now refer to "top" scientists as those who fall in the top 20% of the national rankings per SDS, independently of the university to which they belong. We correlate their numbers, in terms of their occurrence on the total research staff of a university, with the position of the university itself in the rankings at the level of UDA. We then subdivide the rankings into four classes, as provided in the four research profile classes applied by the UK RAE evaluation of universities[16]. Further to this scheme, the Education Funding Council for England (HEFCE) has adopted a PBRF program[17] which does not assign any funds to universities that placed in the bottom class. Universities with an evaluation of their research profile as first class receive (under equal numbers of research staff) three times more funds of universities in the second class, which in turn receive three times as much as those in the third class. By analogy, we divide the university rankings for each Italian UDA into quartiles that simulate a distribution of funds like that of the HEFCE system.

In Table 9, as an example, we present the rankings for the 48 universities active in Physics and, in the last column, the number and percentages of top national scientists in the total Physics research staff of each university.

Clearly, universities placing in the last quartile still include top national scientists among their staff. For example, at the University of Parma, 10 researchers (out of 73 total) have top scientist performance. Udine, last in the Physics rankings, with a research staff of 18 individuals has 3 top scientists. Overall, 62 of the 571 Italian top scientists work in the universities positioned in the last quartile. Under the HEFCE criteria these would be excluded from funding. There are also 204 top national scientists working in the universities that placed in the second quartile, compared to only 156 in the first quartile universities. Paradoxically, a class of universities with a greater number of top scientists would receive one third of the funding given to a class with less top scientists.

A further paradox emerges when we compare the universities in the first and last quartiles. As an example, we consider the situations of the Bari Polytechnic and the University of Salento. The former university, due to its position in the last quartile under an HEFCE type system, would not receive any funds even though it has three national top scientists, representing 21% of its research staff. The University of Salento on the other hand, placing in the first national quartile (12th of 48), would benefit from substantial funding that would support the activities, not only of the 16 top scientists, but also of their 61 institutional colleagues with scientific performance that is less than that of Bari Polytechnic's top scientists.

---

[16] In reality, the RAE includes a fifth level, termed "Unclassified", which is not considered in our simulation. For detail see http://www.rae.ac.uk/aboutus/quality.asp, last access on January 26, 2011.

[17] For detail: http://www.hefce.ac.uk/research/funding/qrfunding/, last access on January 26, 2011.

| rank | University | Top scientists | rank | University | Top scientists |
|---|---|---|---|---|---|
| 1 | University of Brescia | 5 (31%) | 25 | University of Salerno | 7 (15%) |
| 2 | University "G. D'Annunzio" in Chieti | 3 (38%) | 26 | University of Milan | 21 (17%) |
| 3 | Milan Polytechnic | 27 (48%) | 27 | University of Urbino "Carlo Bo" | 3 (33%) |
| 4 | Scuola Normale Superiore of Pisa | 7 (29%) | 28 | University of Bologna | 23 (16%) |
| 5 | University of Trento | 18 (41%) | 29 | University of Firenze | 20 (16%) |
| 6 | International School for Advanced Studies of Trieste | 16 (48%) | 30 | University of Genoa | 14 (14%) |
| 7 | University "Ca' Foscari" in Venice | 3 (43%) | 31 | University of Pisa | 16 (15%) |
| 8 | University of Calabria | 18 (33%) | 32 | University of Catania | 10 (10%) |
| 9 | University of Camerino | 7 (27%) | 33 | University of Rome "Tor Vergata" | 14 (14%) |
| 10 | University of Palermo | 20 (27%) | 34 | University of Basilicata | 2 (17%) |
| 11 | Torino Polytechnic | 16 (34%) | 35 | University of Messina | 7 (13%) |
| 12 | University of Salento | 16 (21%) | 36 | University of Trieste | 12 (18%) |
| 13 | University of Modena and Reggio E. | 14 (33%) | 37 | University of Parma | 10 (14%) |
| 14 | University of Insubria | 9 (32%) | 38 | University of L'Aquila | 10 (19%) |
| 15 | Ancona Polytechnic | 5 (31%) | 39 | University of Perugia | 5 (11%) |
| 16 | University of Milan "Bicocca" | 20 (27%) | 40 | University of Verona | 2 (25%) |
| 17 | University of Naples "Federico II" | 36 (20%) | 41 | University of Bari | 8 (12%) |
| 18 | University of Rome "La Sapienza" | 41 (21%) | 42 | University of Cagliari | 10 (18%) |
| 19 | University of Eastern Piedmont | 2 (12%) | 43 | University of Pavia | 11 (14%) |
| 20 | University of Padua | 32 (20%) | 44 | Sacred Heart Catholic University | 0 (0%) |
| 21 | Second University of Naples | 1 (7%) | 45 | University of Sassari | 0 (0%) |
| 22 | University of "Roma Tre" | 15 (28%) | 46 | Bari Polytechnic | 3 (20%) |
| 23 | University of Ferrara | 10 (20%) | 47 | University of Siena | 0 (0%) |
| 24 | University of Turin | 19 (19%) | 48 | University of Udine | 3 (17%) |

*Table 9: Ranking and subdivision of list into quartiles for universities active in Physics, with numerosity of their top national scientists*

The phenomenon illustrated for Physics is pertinent to practically all UDAs. Table 10 shows the occurrence of top scientists in universities ranked in the last quartile for each UDA. The extreme case of Medicine, the largest area for size, is particularly striking. Here, 490 of the top 2,457 national scientists work in universities where the average performance is in the last national quartile. Overall, 13% of the nation's top scientists (1,045 of 7,980 total), despite top personal scientific performance, would receive no funds under the HEFCE scheme.

| UDA | Top scientists in universities of the last quartile | Total national top scientists | Incidence |
|---|---|---|---|
| Mathematics and computer sciences | 81 | 708 | 11% |
| Physics | 62 | 571 | 11% |
| Chemistry | 61 | 726 | 8% |
| Earth sciences | 21 | 295 | 7% |
| Biology | 138 | 1,165 | 12% |
| Medicine | 490 | 2,457 | 20% |
| Agricultural and veterinary sciences | 55 | 643 | 9% |
| Civil engineering | 28 | 296 | 9% |
| Industrial and information engineering | 109 | 1,110 | 10% |
| Total | 1,045 | 7,980 | 13% |

*Table 10: Occurrence of top national scientists in universities positioned in the last quartile, per UDA*

## 4. Conclusions

The rapid diffusion of national research assessment exercises with the aim of improving allocative efficiency in public funding of individual institutions poses an interesting series of questions for scholars in the field. The literature suggests that permanent adoption of these types of initiatives is desirable, given that their primary objective is to stimulate and reward excellence in public research organizations. However, the debate seems to ignore the importance of the context into which such systems are inserted. Logic suggests that basing government funding on rankings from university research assessment exercises will be more efficient in competitive higher education systems, such as those in the English-speaking nations. The competition in such systems, which long predates the implementation of research assessment exercise, has caused the emergence of top universities with staffs of high average research performance and low variability in performance. However in non-competitive systems, the variability of performance within universities is higher than it is between them. In competitive systems, selective allocation of university funding has a high probability of funding the top scientists, since they are concentrated in the best universities. In less competitive systems, with high dispersion of top scientist among universities, selective funding will inevitably be less efficient, unless assessment exercises also make available individual-level performance rankings to inform selective allocation within universities (for example as proposed in a recent work by Abramo and D'Angelo, 2010). In other words, the macroeconomic aims of research assessment exercises conducted in less competitive higher education systems can only be obtained if the logic of "funds for quality" extends down to all levels of decision-making, particularly to the internal redistribution of the received funds. If this does not occur then it is possible that, lacking incentives for individual scientists, the objective of stimulating better efficiency will not be reached.

To test this logic, the study proposed an analysis of the Italian academic system, of all universities active in the hard sciences over the period 2004-2008. The analysis revealed that 29% of researchers produce 71% of total publications. A full 25% of researchers obtained no citations over the five-year period, and 23% produced 77% of total impact. Similar distributions occur within the individual universities. With such high values of concentration and unproductivity, it is above all the top scientists who determine differences in performance among the universities. We compared the observed situation with a hypothetical scenario that excluded the contribution of the top 20% of scientists for each SDS of each university. The rankings under the two scenarios, although correlated, show the biggest negative shifts for universities with a high concentration of performance. In non-competitive higher education systems the risk is that top scientists of lower-ranked universities will be strongly penalized in terms of funding compared to top-scientists in higher-ranked universities. Additionally there is a high probability that top scientists in lower-ranked universities will receive less funds than lower-performer colleagues, who have the good luck to work in universities with a higher average performance.

In the context of higher education systems with high concentration of performance within universities, we recommend funding allocations based on individual level rankings rather than on university rankings in order to support maximal productivity among the research community.